\documentclass[12pt, preprint]{aastex}

\slugcomment{Accepted for publication in {\it The Astrophysical Journal}}

\shorttitle{Pair cascades in IR radiation fields}
\shortauthors{P. Roustazadeh \& M. B\"ottcher}

\begin{document}

\title{VHE Gamma-Ray Induced Pair Cascades in the Radiation Fields
of Dust Tori of AGN: Application to Cen~A}

\author{P. Roustazadeh and M. B\"ottcher\altaffilmark{1}}

\altaffiltext{1}{Astrophysical Institute, Department of Physics and Astronomy, \\
Ohio University, Athens, OH 45701, USA}

\begin{abstract}
The growing number of extragalactic high-energy (HE, $E > 100$~MeV) 
and very-high-energy (VHE, $E > 100$~GeV) $\gamma$-ray sources that do
not belong to the blazar class suggests that VHE $\gamma$-ray production 
may
be a common property of most radio-loud Active Galactic Nuclei (AGN).
In a
previous paper, we have investigated the signatures of Compton-supported
pair cascades initiated by VHE $\gamma$-ray absorption in monochromatic
radiation fields, dominated by Ly$\alpha$ line emission from the Broad
Line Region. In this paper, we investigate the interaction of nuclear
VHE $\gamma$-rays with the thermal infrared radiation field from a
circumnuclear dust torus. 
Our code follows the spatial development 
of the
cascade in full 3-dimensional geometry. We provide a model fit 
to the
broadband SED of the dust-rich, $\gamma$-ray loud radio galaxy 
Cen~A and show that typical blazar-like jet parameters may be used to model
the broadband SED, if one allows for an additional cascade contribution
to the {\it Fermi} $\gamma$-ray emission.
\end{abstract}
\keywords{galaxies: active --- galaxies: jets --- gamma-rays: galaxies
--- radiation mechanisms: non-thermal --- relativistic processes}

\section{Introduction}

Blazars are a class of radio-loud active galactic nuclei (AGNs) 
comprised of
Flat-Spectrum Radio Quasars (FSRQs) and BL~Lac objects. 
Their spectral energy distributions (SEDs) are characterized
by 
non-thermal continuum spectra with a broad low-frequency component
in the radio -- UV or X-ray frequency range and a high-frequency
component from X-rays to $\gamma$-rays, and they often exhibit
substantial variability across the electromagnetic spectrum. In
the VHE $\gamma$-ray regime, the time scale of this variability 
has been observed to be as short as just a few minutes 
\citep{albert07,aharonian07}. While previous generations of
ground-based Atmospheric Cherenkov Telescope (ACT) facilities 
detected almost exclusively high-frequency peaked BL~Lac objects 
(HBLs) as extragalactic sources of VHE $\gamma$-rays (with the
notable exception of the radio galaxy M87), in recent years, 
a number of non-HBL blazars and even non-blazar radio-loud AGN 
have been detected by the current generation of ACTs. This suggests 
that most blazars might be intrinsically emitters
of VHE $\gamma$-rays. 
According to AGN unification schemes \citep{up95},
radio galaxies are the mis-aligned parent population of blazars with the
less powerful FR~I radio galaxies corresponding to BL~Lac objects and
FR~II radio galaxies being the parent population of radio-loud quasars.
Blazars are those objects which are viewed at a small angle with respect
to the jet axis. If this unification scheme holds, then, by inference,
also radio galaxies may be expected to be intrinsically emitters of VHE
$\gamma$-rays within a narrow cone around the jet axis.

While there is little evidence for dense radiation environments 
in the
nuclear regions of BL~Lac objects --- in particular, HBLs 
---, strong line emission in Flat Spectrum Radio Quasars (FSRQs) 
as well
as the occasional detection of emission lines in the spectra of some
BL~Lac objects \citep[e.g.,][]{vermeulen95} indicates dense nuclear
radiation fields in those objects. This is supported by spectral modeling
of the SEDs of blazars using leptonic models which prefer scenarios
based on external radiation fields as sources for Compton scattering
to produce the high-energy radiation in FSRQs, LBLs and also some
IBLs \citep[e.g.,][]{ghisellini98,madejski99,bb00,acciari08}. If the
VHE $\gamma$-ray emission is indeed produced in the high-radiation-density
environment of the broad line region
(BLR) and/or the dust torus of an
AGN, it is expected to be strongly attenuated by $\gamma\gamma$ pair
production \citep[e.g.][]{pb97,donea03,reimer07,liu08,sb08}. 
\cite{akc08}
have suggested that such intrinsic $\gamma\gamma$ absorption may be
responsible for producing the unexpectedly hard intrinsic (i.e., after
correction for $\gamma\gamma$ absorption by the extragalactic background
light) VHE $\gamma$-ray spectra of some blazars at relatively high redshift. 
A similar effect has been invoked by \cite{ps10} to explain the spectral
breaks in the {\it Fermi} spectra of $\gamma$-ray blazars.
This
absorption process will lead to the development of Compton-supported
pair cascades in the circumnuclear environment \citep[e.g.,][]{bk95,sb10,rb10}.

In \cite{rb10}, we considered the full 3-dimensional development of
a Compton-supported VHE $\gamma$-ray induced cascade in a monochromatic
radiation field. This was considered as an approximation to BLR emission
dominated by a single (e.g., Ly$\alpha$) emission line. In that work,
we showed that for typical radio-loud AGN parameters rather small
($\sim \mu$G) magnetic fields in the central $\sim 1$~pc of the AGN
may lead to efficient isotropization of the cascade emission in the
{\it Fermi} energy range. We applied this idea to fit the {\it Fermi}
$\gamma$-ray emission of the radio galaxy NGC~1275 \citep{abdo09a}
under the plausible assumption that this radio galaxy would appear
as a $\gamma$-ray bright blazar when viewed along the jet axis.

In this paper, we present a generalization of the Monte-Carlo cascade
code developed in \cite{rb10} to arbitrary radiation fields. In particular,
we will focus on thermal blackbody radiation fields, representative of the
emission from a circum-nuclear dust torus. In Section \ref{setup} we will
outline the general model setup and assumptions and describe the modified
Monte-Carlo code that treats the full three-dimensional cascade development.
Numerical results for generic parameters will be presented in Section
\ref{parameterstudy}. In Section \ref{CenA}, we will demonstrate that the
broad-band SED of the radio galaxy Cen~A, including the recent {\it Fermi}
$\gamma$-ray data \citep{abdo09c}, can be modeled with plausible parameters
expected for a mis-aligned blazar, allowing for a contribution from VHE
$\gamma$-ray induced cascades in the {\it Fermi} energy range. We summarize
in Section \ref{summary}.

\begin{figure}[ht]
\centering
\includegraphics[width=15cm]{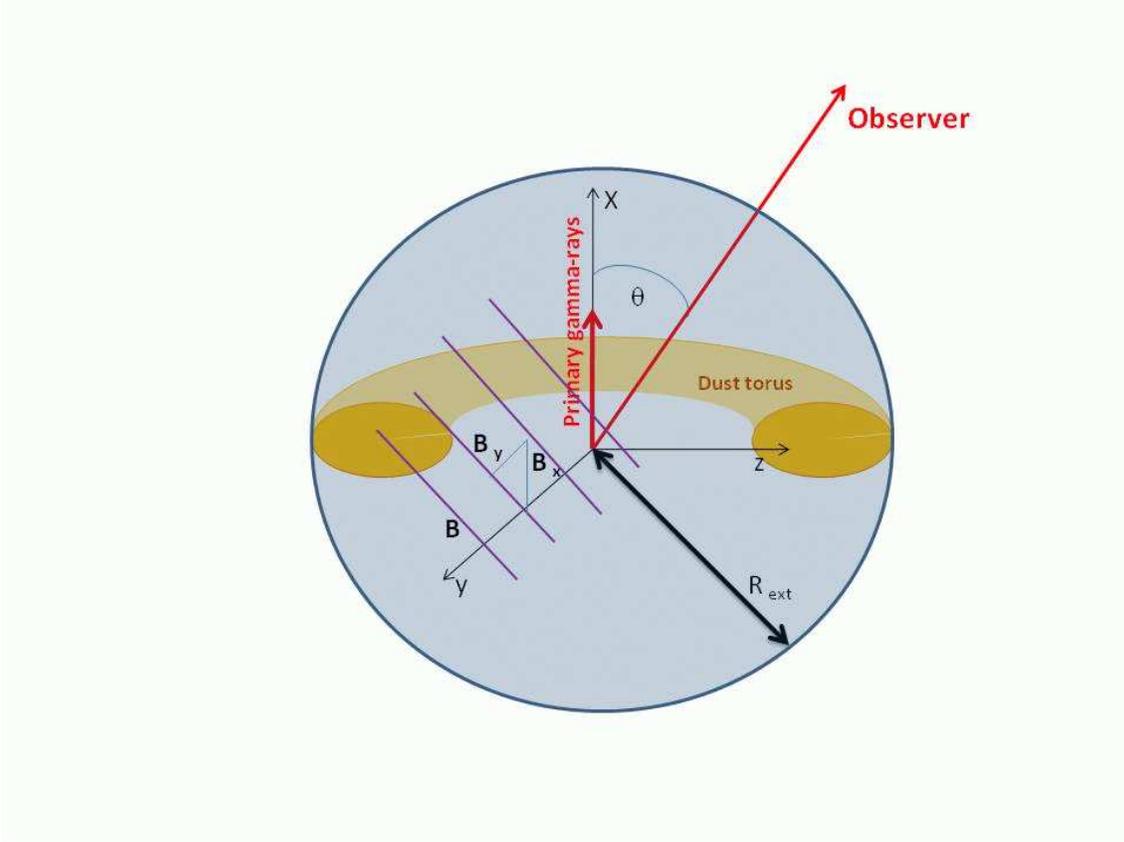}
\caption{\label{geometry}Geometry of the model setup.}
\end{figure}

\section{\label{setup}Model Setup and Code Description}

Figure \ref{geometry} illustrates the geometrical setup of our model system.
We represent the primary VHE $\gamma$-ray emission as a
mono-directional beam of 
$\gamma$-rays propagating along the X axis, described by a power-law with photon 
spectrum index
$\alpha$ and a high-energy cut-off at $E_{\gamma, max}$. For the 
following study, we assume that the primary $\gamma$-rays interact via $\gamma\gamma$
absorption and pair production with an isotropic thermal blackbody radiation field 
within a fixed boundary, given by a radius $R_{\rm ext}$, i.e.,

\begin{equation}
u_{\rm ext} (\nu, r, \Omega) = 2 h \nu^3 / c^3 \frac{A}{\exp(\frac{h\nu}{k T})-1}
\, H(R_{\rm ext} - r)
\label{uapprox}
\end{equation}
where $A$ is a normalization factor chosen to obtain a total radiation energy 
density $u_{\rm ext}$ (see Eq. \ref{uext} below), and
$H$ is the Heaviside function, $H(x) = 1$ if $x > 0$ and $H(x) = 0$ otherwise.
A magnetic field of order $\sim $mG is present. Without loss of generality, we
choose the $y$ and $z$ axes of our coordinate system such that the magnetic field
lies in the (x,y) plane.

The input parameters to our model simulation describing the external radiation 
field are the integral of $u_{\rm ext} (\nu, r, \Omega)$ over all frequencies:

\begin{equation}
u_{\rm ext}=4\pi\int_0 ^\infty u_{\rm ext} (\nu, r, \Omega) d\nu,
\label{uext}
\end{equation}
the blackbody temperature $T$, and the radial extent $R_{\rm ext}$.

We have used the Monte-Carlo code developed by \cite{rb10}. 
This code evaluates $\gamma\gamma$ absorption and pair production using the
full analytical solution to the pair production spectrum of \cite{bs97} under
the assumption that the produced electron and positron travel along the direction
of propagation of the incoming $\gamma$-ray. The trajectories of the particles are 
followed in full 3-D geometry. Compton scattering is evaluated using the head-on 
approximation, assuming that the scattered photon travels along the direction of 
motion of the electron/positron at the time of scattering. While the Compton 
energy loss to the elecron is properly accounted for, we neglect the recoil 
effect on the travel direction of the electron.
In order to improve the statistics of the otherwise very
few highest-energy photons, we introduce a statistical weight, $w$, inversely
proportional to the square of the photon energy, $w = 1/\epsilon^2$. Where 
$\epsilon = \frac{E_{\gamma}}{m_e c^2}$.

To save CPU time, we precalculate tables for the absorption
opacity $\kappa_{\gamma\gamma}$, Compton scattering length $\lambda_{\rm IC}$, 
and Compton cross section for each photon energy, electron energy and interaction
angle before the start of the actual simulation.

The simulation produces a photon event file, logging the energy, statistical
weight, and travel direction of every photon that escapes the region bounded by
$R_{\rm ext}$. A separate post-processing routine is then used to extract 
angle-dependent photon spectra with arbitrary angular and energy binning
from the log files.

\begin{figure}[ht]
\vskip 1.5cm
\centering
\includegraphics[width=12cm]{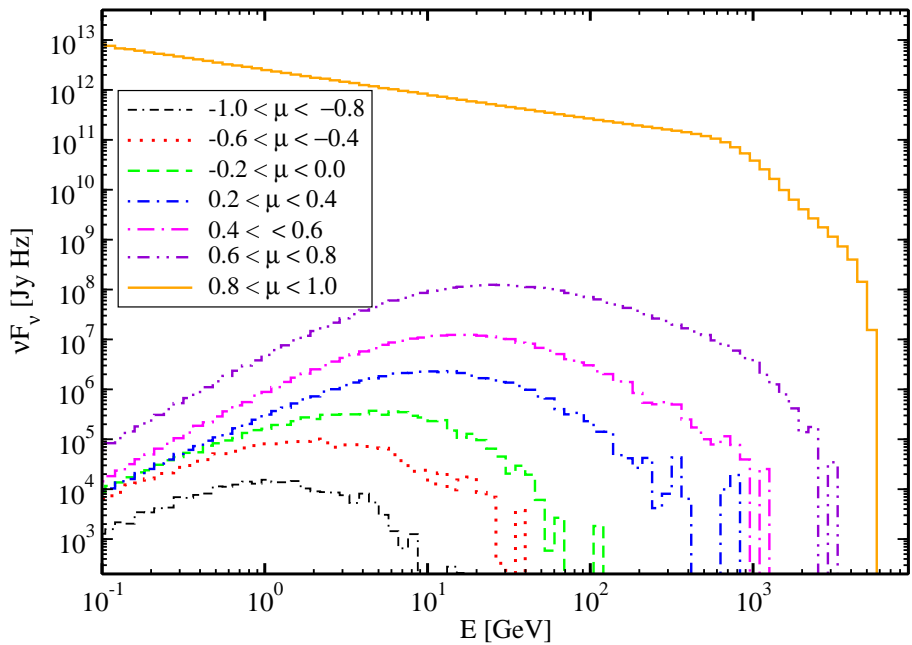}
\caption{\label{standardfig}
Cascade emission at different viewing angles
($\mu = \cos\theta_{\rm obs}$). Parameters: $B = 1 \,$~mG, $\theta_B = 5^o$;
$u_{\rm ext} = 10^{-5}$~erg~cm$^{-3}$, $R_{\rm ext} = 10^{18}$~cm, $T = 1000$~K, 
$\alpha = 2.5$, {$E_{\gamma, {\rm max}} = 5$~TeV}.
}
\end{figure}

\section{\label{parameterstudy}Numerical Results}

For comparison with our previous study on mono-energetic radiation
fields, we conduct a similar parameter study as presented in 
\citep{rb10}, investigating the effects of parameter variations on the
resulting angle-dependent photon spectra. Standard parameters for most
simulations in our parameter study are: a magnetic field of $B = 1$~mG, 
oriented at an angle of $\theta_B = 5^o$
with respect to the X axis 
($B_x = 1$~mG, $B_y = 0.1$~mG); an
external radiation energy density 
of $u_{\rm ext} = 10^{-5}$~erg~cm$^{-3}$, extended
over a region of 
radius $R_{\rm ext} = 10^{18}$~cm; a blackbody temperature of $ T= 10^3$~K
(corresponding to a peak of the blackbody spectrum at a photon energy of 
$E_s^{\rm pk}= 0.25$~eV). The incident $\gamma$-ray spectrum
has a photon 
index of $\alpha = 2.5$
and extends out to $E_{\gamma, {\rm max}} = 5$~TeV.
The emanating photon spectra
for all directions have been normalized with 
the same normalization
factor, corresponding to a characteristic flux level
of a $\gamma$-ray bright ({\it Fermi}) blazar in the forward direction. 

Figure \ref{standardfig} illustrates the viewing angle
dependence of the 
cascade emission. The $\gamma\gamma$ absorption cut-off at an energy 
$E_c = (m_e c^2)^2
/ E_s \sim 1$~TeV is very smooth in this simulation 
because of the broad thermal blackbody spectrum of the external radiation 
filed. In contrast, the $\delta$-function
approximation for the external 
radiation field adopted in \cite{rb10} resulted in an artificially sharp 
cutoff in that work. 

At off-axis viewing angles, the observed spectrum is exclusively the
result of the cascade. In the limit of low photon energies (far below
the $\gamma\gamma$ absorption threshold) and neglecting particle escape
from the cascade zone, one expects
a low-frequency shape close to $\nu F_{\nu} 
\propto \nu^{1/2}$ due to efficient radiative cooling of secondary particles 
injected at high energies \citep{rb10}. However, with the typical parameters
used for this parameter study, the assumption of negligible particle escape
is not always justified. In order to estimate the possible suppression of
the low-frequency cascade emission due to particle escape, we calculate
the critical electron energy for which the Compton cooling time scale
$\tau_{\rm IC} (\gamma)$ equals the escape time scale, $\tau_{\rm esc}
= R_{\rm ext} / (c \, \cos\theta_B)$. Using a characteristic thermal photon
energy of $\epsilon_{\rm Th} = 2.8 \, k T / (m_e c^2)$, the resulting
Compton scattered photon energy, $\epsilon_{\rm esc}$ below which we 
expect to see the effects of particle escape and, hence, inefficient 
radiative cooling, is

\begin{equation}
\epsilon_{\rm esc} = {9 \times 2.8 \, k T \, m_e c^2 \, \cos^2\theta_B \over
16 \, \sigma_T^2 \, u_{\rm ext}^2 \, R_{\rm ext}^2}
\label{epsesc}
\end{equation}
corresponding to an actual energy (in GeV) of
\begin{equation}
E_{\rm esc} \approx 2 \; T_3 \, u_{-5}^{-2} \, R_{18}^{-2} \, \cos^2\theta_B \; {\rm GeV}
\label{Eesc}
\end{equation}
where $T = 10^3 \, T_3$~K, $u_{\rm ext} = 10^{-5} \, u_{-5}$~erg~cm$^{-3}$, and
$R = 10^{18} \, R_{18}$~cm. Therefore, for our standard parameters, we expect
the low-frequency ($E \lesssim$~a few GeV) to be significantly affected by
particle escape. This explains why the low-energy photon spectra shown
in Figure \ref{standardfig} are harder than $\nu^{1/2}$. The cascade emission is 
progressively suppressed at high energies with
increasing viewing angle due to 
incomplete isotropization of the highest-energy secondary
particles. This effect
becomes important beyond the energy $E_{\rm IC, br}$ of Compton-scattered photons 
by secondary electrons/positrons for which the Compton-scattering length $\lambda_{IC}$ 
equals the distance travelled along the gyrational motion over an angle $\theta$, 
$\lambda_{IC}(\gamma) = \theta r_{\rm gyr}(\gamma)$, which is given by 

\begin{equation}
E_{\rm IC, br} = {3 \, e \, B \over 4 \, \sigma_T \, u_{\rm ext} \, \theta}
\, E_s \; \sim \;
1.3 \; B_{-3} \, u_{-5}^{-1} \, T_3 \, \theta^{-1} \; {\rm GeV}.
\label{ICbreak}
\end{equation}

where $B = 10^{-3}$~mG \citep{rb10}.

\begin{figure}[ht]
\vskip 1.5cm
\centering
\includegraphics[width=12cm]{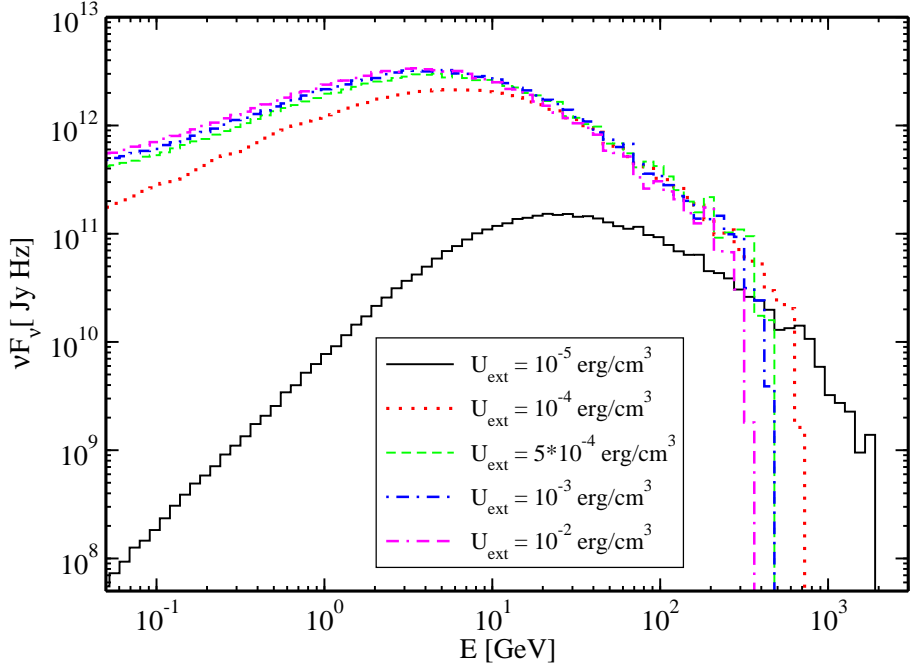}
\caption{\label{ufig}The effect of a varying external radiation energy density.
Parameters: $B_x = 10^{-3}$~G, $B_y = 1.3\times 10^{-4}$ G, $\theta_B = 7.4^0$;
and other parameters are the same as for Figure
\ref{standardfig} in the angular bin $0.4\leq\mu\leq0.6$ }
\end{figure}

Figure \ref{ufig} shows the cascade spectra for different values
of the external radiation field
energy density $u_{\rm ext}$. For large
energy densities $u_{\rm ext} \gtrsim 10^{-4}$~erg~cm$^{-3}$,
$\tau_{\gamma\gamma}
\gg 1$ for photons above the pair production threshold $\gamma\gamma$ so
that essentially all VHE photons will be absorbed and the photon flux from 
the cascade
becomes independent of $u_{\rm ext}$. 

The figure confirms our discussion concerning escape and hence inefficient 
radiative cooling of low-energy particles above (see Eq. \ref{Eesc}). For
large values of $u_{\rm ext}$, the Compton loss time scale for all relativistic
electrons producing $\gamma$-rays in the considered range, is much shorter
than the escape time scale. Hence, the expected $\nu F_{\nu} \propto \nu^{1/2}$
shape results. In the low-$u_{\rm ext}$ case, escape affects even ultrarelativistic 
electrons, resulting in a substantial hardening of the low-energy photon spectrum.

\begin{figure}[ht]
\vskip 1.5cm
\centering
\includegraphics[width=12cm]{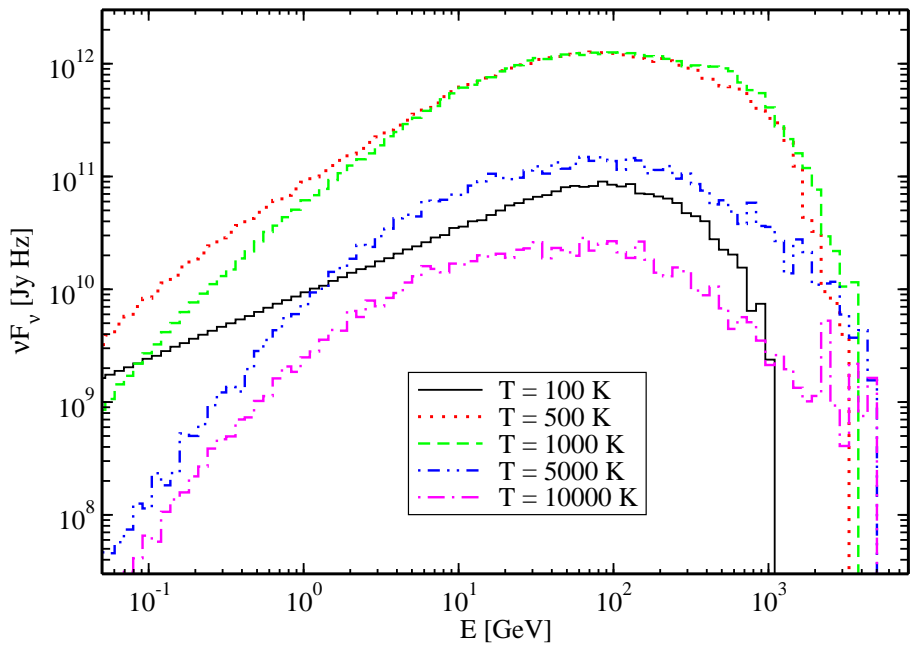}
\caption{\label{Tfig}The effect of a varying temperature of the external
blackbody radiation field. $\theta_B = 15^o$ and other parameters are the 
same as for Figure
\ref{standardfig} in the angular bin $0.4\leq\mu\leq0.6$ }
\end{figure}

Figure \ref{Tfig} illustrates the effect of a varying temperature of the 
external blackbody radiation field. As the temperature increases up to 
$1000$~K the cascade flux increases because the $\gamma\gamma$ absorption 
threshold energy decreases so that an increasing fraction of $\gamma$-rays 
can be absorbed. The isotropization turnover is almost independent of $T$. 
For temperatures $ T > 1000$~K the cascade flux decreases with increasing
$T$ because $u_{\rm ext}$ remains fixed, leading to a decreasing photon 
number density and absorption
opacity $\kappa_{\gamma\gamma}$ with increasing 
$T$ (and, hence, increasing $E_s$). 
The figure also confirms our
expectation (Eq. \ref{Eesc}) that an increasing blackbody temperature
leads to an increasing suppression of the low-frequency portion of the
cascade emission due to particle escape.

\begin{figure}[ht]
\vskip 1.5cm
\centering
\includegraphics[width=12cm]{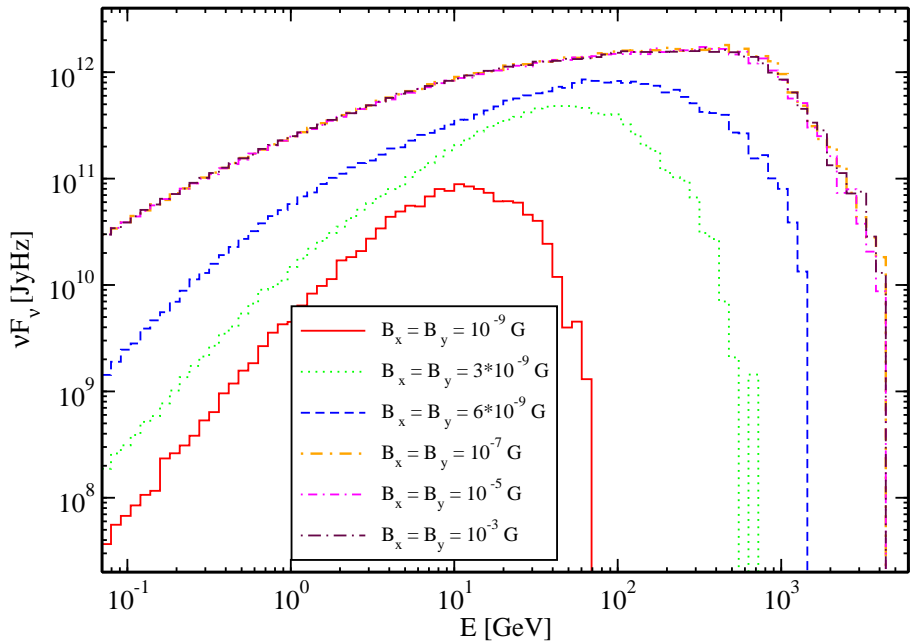}
\caption{\label{Bfig}The effect of a varying magnetic field strength, for a
fixed angle of $\theta_B = 45^o$ between jet axis and magnetic field. All
other parameters are the same as for Figure \ref{standardfig} in the angular
bin $0.4\leq\mu\leq0.6$}
\end{figure}

\begin{figure}[ht]
\vskip 1.5cm
\centering
\includegraphics[width=12cm]{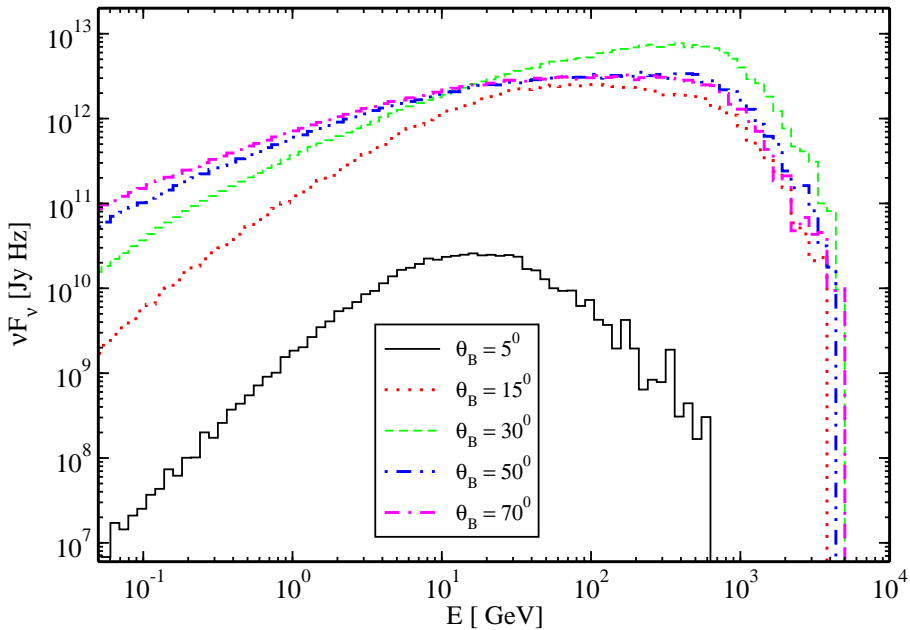}
\caption{\label{Bthetafig}The effect of a varying magnetic field orientation,
for a fixed magnetic field strength of $B = 1$~mG. All other parameters
are the same as for Figure \ref{standardfig} in the angular bin $0.4\leq\mu\leq0.6$}
\end{figure}

Figures \ref{Bfig} and \ref{Bthetafig} illustrate the effects of
varying magnetic-field parameters (strength and orientation). As expected,
the results are essentially the same as for cascades in monoenergetic
radiation fields investigated in \cite{rb10}: The cascade development 
is extremely sensitive to the
transverse magnetic field $B_y$. The  
limit in which even the highest-energy
secondary particles are effectively 
isotropized before undergoing the
first Compton scattering interaction,
is easily reached for typical magnetic fields expected in the circum-nuclear
environment of AGNs. 

\section{\label{CenA}Application to Cen~A}

The standard AGN unification scheme \citep{up95} proposes that blazars
and radio galaxies are intrinsically identical objects viewed at different angles
with respect to the jet axis. According to this scheme, FR~I and FR~II radio galaxies
are believed to be the parent population of BL~Lac objects and FSRQs, respectively.
Hence, if most blazars, including LBLs and FSRQs, are intrinsically VHE $\gamma$-ray
emitters potentially producing pair cascades in their immediate environments, the
radiative signatures of these cascades might be observable in many radio galaxies.
In fact, {\it EGRET} provided evidence for $> 100$~MeV $\gamma$-ray emission from
three radio galaxies (Cen~A: \cite{sreekumar99}, 3C~111: \cite{hartman08}, and
NGC~6251: \cite{muk02}). These sources have been confirmed as high-energy
$\gamma$-ray sources by {\it Fermi} \citep{abdo09c,abdo10b}, along with the
detection of five more radio galaxies (NGC~1275: \cite{abdo09a}, M~87: \cite{abdo09b},
3C~120, 3C~207, and 3C~380: \cite{abdo10b}). In this paper, we focus on the radio
galaxy Cen~A \citep{abdo09c}.

The FR I Cen A is the nearest radio-loud active galaxy to Earth. It has a redshift 
of z=0.00183 at the distance of $D = 3.7 $~Mpc. Recently, the Auger collaboration 
reported that the arrival directions of the highest energy cosmic rays 
($E \gtrsim 6 \times10^{19}$~eV) observed by the Auger observatory are 
correlated with nearby
AGN, including Cen A \citep{abraham07,abraham08}. 
This suggests that Cen A may be the dominant source of observed UHECR 
nuclei above the GZK cut off.

\begin{figure}[ht]
\centering
\includegraphics[width=10cm]{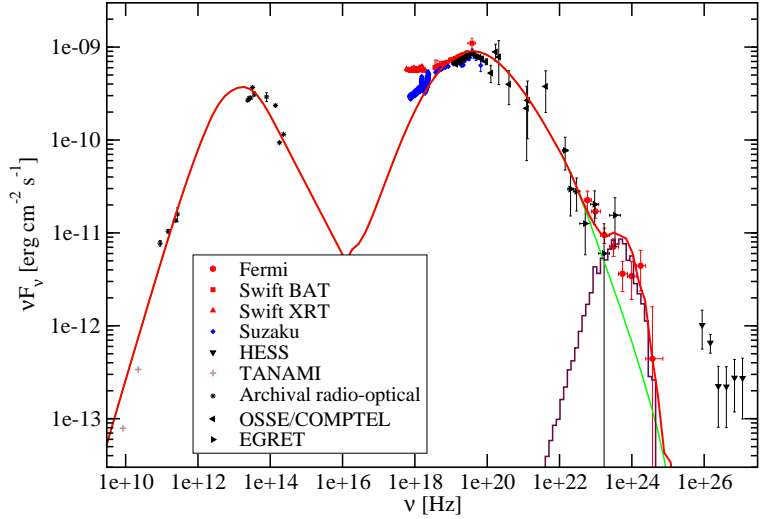}
\caption{\label{fit}Fit to the SED of Cen~A. The green curve is a fit to the
broad-band SED using the model of \citep{bc02}, while the maroon curve is
the cascade emission resulting from $\gamma\gamma$ absorption of the forward
jet emission. The red curve is the sum of both contributions (viewed at an
angle of $70^o$. }
\end{figure}

Cen A has an interesting radio structure on several size scales. The most 
prominent features are its giant radio lobes, which subtend $\sim 10^0$ on 
the sky, oriented primarily in the north-south direction. They have been 
imaged at $4.8$ GHz by the Parkes telescope \citep{junkes93} and studied 
at up to $\sim 60$~GHz by \cite{hardcastle09}. The radio lobes are the
only extragalactic source structure that has so far been spatially resolved 
in GeV $\gamma$-rays by $\emph{Fermi}$ \citep{abdo10c}. The innermost region 
of Cen A has been resolved with VLBI and shown to have a size of 
$\sim 3 \times 10^{16}$ cm \citep{kellermann97,horiuchi06}. Observations at 
shorter wavelengths also reveal a small core. VLT infrared interferometry 
resolves the core size to $\sim 6 \times 10^{17}$~cm \citep{meisen07}.
The angle of the sub-parsec jet of Cen A to our line of sight is 
$\thicksim 50^0-80^0$ \citep{Tingay98} with a preferred value of
$\sim 70^o$ \citep{steinle09}.

The K-band nuclear flux with starlight subtracted is $F(K) \thicksim 38$ mJy, 
corresponding to $\nu L_{\nu} \thicksim 7 \times 10^{40}$~erg~$s^{-1}$ for a 
distance of $3.5$~Mpc \citep{marconi00}. The mid-IR flux of $\thicksim 1.6$~Jy 
at $11.7 \, \mu$m corresponds to $\nu L_\nu \thicksim 6 \times 10^{41}$~erg~$s^{-1}$. 
The broadband SED from radio to $\gamma$-rays has been fitted with a synchrotron 
self-Compton model by \cite{abdo10a}. In their model \citep[see Table 2 of ][]{abdo10a}, 
a maximum electron Lorentz factor of $\gamma_{max}= 1\times10^8$ was required in
order to produce the observed $\gamma$-ray emission. However, given the assumed 
magnetic field of $B=6.2$~G, this does not seem possible since for $\gamma = 10^8$ 
electrons, the synchrotron loss time scale is shorter than their gyro-timescale, 
which sets the shortest plausible acceleration time scale.
Here, we therefore 
present an alternative interpretation of the SED, based on the plausible assumption
that Cen~A would appear as a VHE $\gamma$-ray emitting blazar in the jet direction,
and cascading of VHE $\gamma$-rays on the nuclear infrared radiation field produces
an observable off-axis $\gamma$-ray signature in the {\it Fermi} energy range. 

Figure \ref{fit} illustrates a broadband fit to the SED of Cen A
\citep[data from][]{abdo10a}, using the equilibrium version
of the 
blazar radiation transfer code of \cite{bc02}, as
described in more 
detail in \cite{acciari09}. For this fit,
standard blazar jet 
parameters were adopted, but the viewing
angle was chosen in accord 
with the observationally inferred
range. Specifically, we chose 
$\theta_{\rm obs} = 70^o$. Other model parameters
include a bulk Lorentz factor of $\Gamma = 5$, a radius of
the emission region of $R = 1 \times 10^{16}$~cm, a kinetic
luminosity in relativistic electrons, $L_e = 9.4 \times
10^{43}$~erg~s$^{-1}$, a co-moving magnetic field of $B = 11$~G,
corresponding to a luminosity in the magnetic field (Poynting
flux) of $L_B = 1.1 \times 10^{45}$~erg~s$^{-1}$ and a magnetic-field
equipartition fraction $\epsilon_B \equiv L_B / L_e = 12$, 
corresponding to a Poynting-flux dominated jet. 
Electrons are injected into the emission region at a 
steady rate, with a distribution characterized by low- and 
high-energy
cutoffs at $\gamma_1 = 1.2 \times 10^3$ and $\gamma_2 
= 1.0
\times 10^6$, respectively, and a spectral index of $q = 3.5$.
The code finds a self-consistent equilibrium between particle
injection, radiative cooling and escape, from which the final
photon spectrum is calculated.
The resulting broadband SED fit is illustrated by the solid
green curve in Figure \ref{fit}. The flux emanating in the
forward direction ($\theta_{\rm obs} = 0^o$, i.e., the
blazar direction) has been chosen as an input to our
cascade simulation to evaluate the cascade emission in the
nuclear infrared radiation field of Cen~A observed at the
given angle of $\theta_{\rm obs} = 70^o$.

For the cascade simulation, we assumed a blackbody temperature of 
$2300$~K resulting in a peak frequency in the K-band. The external 
radiation field is
parameterized through $u_{\rm ext} = 1.5 \times 10^{-3}$~erg~cm$^{-3}$ 
and
$R_{\rm ext} = 3 \times 10^{16}$~cm. These parameters combine
to an IR luminosity of $L_{\rm BLR} = 4 \pi R_{\rm ext}^2 \, c \, u_{\rm ext}
= 5 \times 10^{41}$~erg~s$^{-1}$, in agreement with mid-IR flux observed for
Cen A \citep{marconi00}. The magnetic field is $B = 1$~mG, oriented at an 
angle of
$\theta_B = 4^o$. The cascade spectrum shown in Figure \ref{fit} pertains
to the angular bin $0.28 < \mu < 0.38$ (corresponding to $67^o \lesssim
\theta \lesssim 73^o$), appropriate for the known orientation of Cen~A 
and consistent with our broadband SED fit parameters.
The cascade spectrum
is shown by the maroon curve in Figure \ref{fit}, while the total observed
spectrum is the solid red curve. The figure illustrates that the cascade
contribution in the {\it Fermi} range substantially improves the fit,
while still allowing physically reasonable parameters for the broadband
SED fit.

\section{\label{summary}Summary}

We investigated the signatures of Compton-supported pair cascades 
initiated by the interaction of nuclear
VHE $\gamma$-rays with the 
thermal infrared radiation field of a
circumnuclear dust torus in 
AGNs. We follow the spatial development of the
cascade in full 
3-dimensional geometry and study the dependence of the radiative
output on various parameters pertaining to the infrared radiation
field and the magnetic field in the cascade region. 

We confirm the results of our previous study of cascades in monoenergetic
radiation fields that small ($\gtrsim \mu$G) perpendicular (to the primary
VHE $\gamma$-ray beam) magnetic field components lead to efficient
isotropization of the cascade emission out to HE $\gamma$-ray energies. 
The cascade intensity as well as the location of a high-energy turnover
due to inefficient isotropization also depend sensitively
on the energy density and temperature of the soft blackbody radiation
field, as long as the cascade is not saturated in the sense that not
all VHE $\gamma$-rays are absorbed. 

The shape of the low-frequency tail of the cascade emission is
a result of the interplay between radiative cooling and escape. For
environments characterized by efficient radiative cooling, the canonical
$\nu F_{\nu} \propto \nu^{1/2}$ spectrum results. If radiative cooling 
is inefficient compared to escape, the low-frequency cascade spectra 
are harder than $\nu^{1/2}$. 

We provide a model fit to the
broadband SED of the dust-rich, $\gamma$-ray 
loud radio galaxy Cen~A.
We show that typical blazar-like jet parameters 
may be used to model
the broadband SED, if one allows for an additional 
cascade contribution
to the {\it Fermi} $\gamma$-ray emission due to
$\gamma\gamma$ absorption and cascading in the thermal infrared radiation
field of the prominent dust emission known to be present in Cen~A.

\acknowledgements{We thank J. Finke for providing the SED data points of
Cen~A. This work was supported by NASA through Fermi Guest Investigator
Grants NNX09AT81G and NNX10AO49G.}


\begin{thebibliography}{}

\bibitem[Abdo et al.(2009a)]{abdo09a}Abdo, A. A., et al., 2009a, ApJ, 699, 31
\bibitem[Abdo et al.(2009b)]{abdo09b}Abdo, A. A., et al., 2009b, ApJ, 707, 55
\bibitem[Abdo et al.(2009c)]{abdo09c}Abdo, A. A., et al., 2009c, ApJ, 700, 597
\bibitem[Abdo et al.(2010a)]{abdo10a}Abdo, A. A., et al., 2010a, ApJ, 719, 1433
\bibitem[Abdo et al.(2010b)]{abdo10b}Abdo, A. A., et al., 2010b, ApJ, 720, 912
\bibitem[Abdo et al.(2010c)]{abdo10c}Abdo, A. A., et al., 2010c, Science, 328, 725
\bibitem[Abraham, J., et al. (2007)]{abraham07}Abraham, J., et al., 2007, Science, 318, 938
\bibitem[Abraham, J., et al. (2008)]{abraham08}Abraham, J., et al., 2008, Astropar. Phys., 29, 188
\bibitem[Acciari et al.(2008)]{acciari08}Acciari, V. A., et al., 2008, ApJ, 684, L73
\bibitem[Acciari et al.(2009)]{acciari09}Acciari, V. A., et al., 2009, ApJ, 707, 612
\bibitem[Aharonian et al.(2007)]{aharonian07}Aharonian, F., et al., 2007, ApJ, 664, L71
\bibitem[Aharonian et al.(2008)]{akc08}Aharonian, F. A., Khangulyan, D., \& Costamante, L.,
2008, MNRAS, 387, 1206
\bibitem[Albert et al.(2007)]{albert07}Albert, J., et al., 2007, ApJ, 669, 862

\bibitem[Bednarek \& Kirk(1995)]{bk95}Bednarek, W., \& Kirk, J. G., 1995, A\&A, 294, 366
\bibitem[B\"ottcher \& Chiang(2002)]{bc02}B\"ottcher, M., \& Chiang, J., 2002, ApJ, 581, 127
\bibitem[B\"ottcher \& Bloom(2000)]{bb00}B\"ottcher, M., \& Bloom, S. D., 2000, AJ,
119, 469
\bibitem[B\"ottcher \& Schlickeiser(1997)]{bs97}B\"ottcher, M., \& Schlickeiser, R.,
1997, A\&A, 325, 866
\bibitem[B\"ottcher(2007)]{boettcher07}B\"ottcher, M., 2007, in proc. ``The Multimessenger
Approach to Gamma-Ray Sources'', ApSS, 309, 95
\bibitem[Dermer \& B\"ottcher(2006)]{db06}Dermer, C. D., \& B\"ottcher, M., 2006,
ApJ, 643, 1081
\bibitem[Donea \& Protheroe(2003)]{donea03}Donea, A. C., \& Protheroe, R. J., 2003,
Astrop. Phys., 18, 337

\bibitem[Ghisellini et al.(1998)]{ghisellini98}Ghisellini, G., et al., 1998, MNRAS,
301, 451

\bibitem[Hartman et al.(2008)]{hartman08}Hartman, R. C., Kadler, M., \& Tueller, J.,
2008, ApJ, 688, 852
\bibitem[Hardcastle et al.(2009)]{hardcastle09}Hardcastle, M. J., Cheung, C. C., 
Feain, I. J., \& Stawarz, {\L}.2009, MNRAS, 393, 1041

\bibitem[Horiuchi et al.(2006)]{horiuchi06}Horiuchi, S., Meier, D. L., Preston, R. A.,
\& Tingay, S. J., 2006, PASJ, 58, 211

\bibitem[Junkes et al.(1993)]{junkes93}Junkes, N., Haynes, R. F., Harnett, J. I., 
\& Jauncy, D. L., 1993, A\&A, 269, 29

\bibitem[Kellermann et al.(1997)]{kellermann97}Kellerman, K. I., Zensus, J. A., \& Cohen, M. H.,
1997, ApJ, 475, L93

\bibitem[Liu et al.(2008)]{liu08}Liu, H. T., Bai, J. M., \& Ma, L., 2008, ApJ, 688, 148

\bibitem[Madejski et al.(1999)]{madejski99}Madejski, G. M., et al., 1999, ApJ, 521, 145
\bibitem[Marconi et al. (2000)]{marconi00}Marconi, A., Schreire, E. J., Koekemoer, A., 
Capetti, A., Axon, D., Maccetto, D., \& Caon, N. 2000, ApJ, 528, 276
\bibitem[Meisenheimer et al.(2007)]{meisen07}Meisenheimer, K., et al., 2007, 
A\& A, 471, 453
\bibitem[Mukherjee et al.(2002)]{muk02}Mukherjee, R., Halpern, J., Mirabal, N., \&
Gotthelf, E. V., 2002, ApJ, 574, 693

\bibitem[Poutanen \& Stern(2010)]{ps10}Poutanen, J., \& Stern, B., 2010, ApJ, 717, L118
\bibitem[Protheroe(1986)]{protheroe86}Protheroe, R. J., 1986, MNRAS, 221, 769
\bibitem[Protheroe \& Biermann(1997)]{pb97}Protheroe, R. J., \& Biermann, P. L., 1997,
Astrop. Phys., 6, 293

\bibitem[Reimer(2007)]{reimer07}Reimer, A., 2007, ApJ, 665, 1023
\bibitem[Roustazadeh \& B\"ottcher(2010)]{rb10}Roustazadeh, P., \& B\"ottcher, M.,
2010, ApJ, 717, 468

\bibitem[Sitarek \& Bednarek(2008)]{sb08}Sitarek, J., \& Bednarek, W., 2008, MNRAS, 391, 624
\bibitem[Sitarek \& Bednarek(2010)]{sb10}Sitarek, J., \& Bednarek, W., 2010, MNRAS, 401, 1983
\bibitem[Sreekumar et al.(1999)]{sreekumar99}Sreekumar, P., Bertsch, D. L., Hartman, R. C.,
Nolan, P. L., \& Thompson, D. J., 1999, Astropart. Phys., 11, 221
\bibitem[Steinle(2009)]{steinle09}Steinle, H., 2009, in proc. of ``The Many Faces of
Centaurus A'', PASA, in press
\bibitem[Tingay et al.(1998)]{Tingay98}Tingay, S. J., et al., 1998, AJ, 115, 960
\bibitem[Urry \& Padovani(1995)]{up95}Urry, C. M., \& Padovani, P., 1995, PASP, 107, 803

\bibitem[Vermeulen et al.(1995)]{vermeulen95}Vermeulen, R. C., et al., 1995, ApJ, 452, L5


\bibitem[Zdziarski(1988)]{zdziarski88}Zdziarski, A. A., 1988, ApJ, 335, 786

\end{thebibliography}
\end{document}